\documentclass[letter]{aa}

\usepackage[varg]{txfonts}
\usepackage{graphicx}
\usepackage{hyperref}
\usepackage{booktabs}
\usepackage{subcaption}
\usepackage{natbib}
\bibpunct{(}{)}{;}{a}{}{,} 

\newcommand\new[1]{\textcolor{black}{#1}}

\begin{document}

\title{Asteroid (4337) Arecibo: Two ice-rich bodies forming a binary }

\subtitle{Based on \textit{Gaia} astrometric data}

\author{Ziyu Liu\inst{1} \and  Daniel Hestroffer \inst{1}
    \and Josselin Desmars\inst{2,1} \and Pedro David\inst{1}
  }
\institute{IMCCE, Paris Observatory, CNRS, univ. PSL, Sorbonne université, univ. Lille, 77 Av. Denfert-Rochereau, 75014 Paris, France
              \email{ziyu.liu@obspm.fr; daniel.hestroffer@obspm.fr}
         \and 
         Institut Polytechnique des Sciences Avancées IPSA, 63b Bd. de Brandebourg, 94200, Ivry-sur-Seine, France \\
             }

\date{Received 2 May 2024 / Accepted 18 July 2024}

\abstract
   {Binary asteroids are present in all populations of the Solar System, from near-Earth to trans-Neptunian regions. As is true for the small Solar System bodies (SSSBs), binary asteroids generally offer valuable insights into the formation of the Solar System, as well as its collisions and dynamic evolution. In particular, the binaries provide fundamental quantities and properties of these SSSBs, such as mass, angular momentum, and density, all of which are often hidden. The direct measurement of densities and porosities is of great value in revealing the gravitational aggregates and icy bodies that form the asteroid-comet continuum.}
   {Several observation techniques from space and ground-based platforms have provided many results in this regard. Here we show the value of the \textit{Gaia} mission and its high-precision astrometry for   analysing  asteroid binaries \new{and for  individually deriving the masses of the  components.} }
   {We focus on the binary asteroid (4337) Arecibo, a member of the Themis family. We analysed the astrometry obtained in the \textit{Gaia} FPR catalogue release, and performed orbital fitting for both the heliocentric orbit of the system and the relative orbit of the binary components.}
   {We obtain an estimation of the component masses and their flux ratio, and derive bulk densities $\rho_1\approx1.2$ and $\rho_2\approx1.6$ for the primary and the secondary, respectively. The results are consistent with an ice-rich body in the outer main belt. They also show a significantly denser secondary or a less closely packed primary. 
   Constraints on these densities and on macroscopic porosities are nevertheless limited by our poor knowledge of the sizes of the  components. Observations of future mutual events, and of stellar occultations predicted in 2024--2025, will be essential for improving our knowledge of this system and its formation.}
    {}
\keywords{Gaia, Binary Asteroids, (4337) Arecibo}
\maketitle

%
\section{Introduction}
\textit{Gaia} is a space mission of the European Space Agency (ESA) that was launched in 2013. The mission opens a window to explore the unprecedented high-precision astrometric data for a large population of asteroids. Its latest data release,  the Focused Product Release (\textit{Gaia} FPR) published in October 2023,  contains 66 months of data for about 160,000 asteroids. By covering a main-belt asteroid's typical orbital period, it has been shown that the \textit{Gaia} data alone can provide very precise heliocentric orbits \citep{david23}. As shown in previous data releases, the \textit{Gaia} data is precise along its scanning direction (the AL axis), in contrast to the perpendicular across-scan direction. In addition, the standard deviation at the CCD level is of the order of 0.2\,mas (milliarcsecond) for an object of magnitude $G \approx 13$ \citep{david23}. As a result, all data analyses and comparisons are done on this AL axis in the rest of this Letter.

\new{Of more than 500 known binary and multiple systems in the Solar System, to date \textit{Gaia} has  observed approximately 300 of them in different regions. }Thanks to the unprecedented precision of \textit{Gaia}, it is possible to reveal the astrometric signature of binary asteroids. This is the case for the recently discovered binary (4337) Arecibo system (hereafter Arecibo), analysed with the \textit{Gaia} DR3 data \citep{tanga23}, where the astrometric wobble was clearly detected in a time window of several days covering successive transits (see their Fig.~15 and Sect.~5.2.1). 

In the present work we report on our continuing research on the Arecibo binary system by taking into account all FPR observations, corresponding to a total of 691 observations, or 90 transits. We performed a fit on both the heliocentric orbit and the relative orbit of the binary over approximately five\,years of data. By adding the positions issued from two stellar occultations \citep{2022MPBu...49....3G}, we were able to constrain the orbital and physical parameters of the binary. We thus derived the complete relative orbit, the flux ratio, and the mass of each component, and also give an estimate of their bulk densities and the corresponding ratio.


The second section presents the method and data model implemented. The third and forth sections are dedicated to the results and discussions.

\section{Method}
\subsection{Heliocentric orbit}
\label{S:helorb}
The heliocentric orbit is fitted to the FPR data by making use of the Numerical Integration of the Motion of Asteroids (NIMA) programme, presented in \citet{2015A&A...575A..53D}. The orbit propagation and the computation of the partial derivatives  includes the planetary ephemerides with four additional massive asteroids as perturbers, as well as relativistic corrections (on the observations and for the equation of motion). NIMA applies the weighted least-squares method to compute a differential correction to the initial state vector fitting the observations. The weighting scheme follows the \textit{Gaia} error model for systematic and random error provided in the \textit{Gaia} catalogue \citep[chapter 8]{gdr3-22}. At this stage the post-fit residuals reflect, \new{among} other things, systematic effects due to the offset between the system's computed centre of mass (CoM) and the observed centre of light (photocentre).

\subsection{Binary photocentre--barycentre offset}
In the case of a binary system, the total photocentre displacement is the result of the offset due to the presence of the satellite, added to the offsets from the components' apparent discs and solar phase angle. Following \citet{2012P&SS...73...56P}, this offset $\delta\eta$ along AL can be written as
\begin{equation}\label{eq:r}
    \delta\eta = \left(\frac{I_2}{I} - \frac{M_2}{M}\right)\,\Delta\eta + \delta r_1 + \delta r_2
,\end{equation} 
where $\Delta\eta$ is the angular separation between the primary and the secondary projected onto the AL direction; $I=I_1+I_2$ and $M=M_1+M_2$ are the total intensity and total mass, respectively; and $O_p = \delta r_1 + \delta r_2$ is the projected photocentre offset due to the apparent discs of the components. This last term $O_p$ can be neglected in a first approximation. 

From the point of view of the \textit{Gaia} telescope and its separation power, we had to consider different observational cases where a binary system can be either resolved or unresolved. An analytical model has been proposed in \citet{LL:LL-136} to describe the photocentre versus CoM shift ($\delta\eta$) as function of the  flux ratio ($f = I_2/I_1$), and mass ratio ($q = M_2/M_1$) of the binary components:
\begin{equation}\label{eq:cases}
  \delta\eta =
    \begin{cases}
      \left(\frac{f}{1+f} -\frac{q}{1+q} \right) \Delta\eta & \text{if $ \left| \Delta\eta/u \right| \leq 0.1 $ }\\
      u\,B(f,\Delta\eta/u) - \frac{q}{1+q} \Delta\eta & \text{if $ 0.1 < \left|\Delta\eta/u \right| \leq 3-f$}\\
      -\frac{q}{1+q} \Delta\eta & \text{if $ 3-f < \left|\Delta\eta/u \right|$}\\
    \end{cases}       
\end{equation}
Here $u = 90\,$mas is the `resolution unit' of \textit{Gaia} (3/2 pixel size in AL). 

\noindent 
In the first case, the binary system is not resolved. The recorded position corresponds to the binary's centre of light, and the offset refers to the difference between the CoM and the photocentre of the whole system. Under the approximation of two spherical bodies, of the same albedo and the same density, this unresolved astrometric wobble amplitude can be expressed as function of the mass ratio alone: $w = \left(1+q^{-2/3}\right)^{-1} -  \left(1+q^{-1}\right)^{-1}$.\\
The second case is considered partially resolved. In this case the difference between the CoM and the photocentre depends on 
the bias function $B(f,\Delta\eta/u)$, which is the root of the equation $x + (x-p)f\exp{(p.x-p^2/2)} = 0$ \citep[]{LL:LL-136}. The bias is linked to the flux difference $f$ and scaled projected separation $p=\Delta\eta/u$. \\
In the third case the system is resolved, and the offset corresponds to the position of the primary body with respect to the system's centre of mass (assuming that only the primary is observed).

\subsection{Relative orbit}
We also developed a program to compute the Cartesian coordinates of the satellite with respect to the primary. This was done, under the assumption of  two-body Keplerian motion, by taking into account the complete relative orbit (with seven orbital parameters: semi-major axis $a$, longitude of ascending node $\Omega$, inclination $i$, argument of pericentre $\omega$, eccentricity $e$, pericentre passed time $t0$, and period $T$), the flux ratio $f$, and the mass ratio $q$. This provides the separation $\Delta\eta$ projected along the AL direction to be compared to the residuals from Sect.~\ref{S:helorb}.


For each transit we computed a normal point by taking the mean AL residuals $\bar x$ for the $N$ CCDs, and an associated standard deviation $\sigma_{\bar{x}} = {\sqrt{\sum^N_{i=1} \sigma^2_i}}/(N-1)$.
We put a three-sigma threshold on these normal points to remove the  lower quality data and the outliers. This resulted in a final set of  89 data points to perform the relative orbit fitting at transit level. This orbit fitting was done by making use of a Markov chain Monte Carlo (MCMC) method, 
minimizing the weighted mean square error $(\bar{x}-\delta\eta)/\sigma_{\bar{x}}$ on the offset in Eq.~(\ref{eq:cases}).
The initial value was obtained from a simplex fit of the data; the prior followed a uniform distribution. A few hundred burn-in steps and a few thousand iterations were performed for each experiment. 

As  expected, we found that a strong correlation exists between the component masses and their flux ratio; we thus introduced a correcting factor $s$ such that
    $q = s.f^{3/2}$, 
and eventually fit for $f$ and $s$ as free parameters. We also note that if both binary components are spherical bodies, with the same density and the same albedo, then $s=1$. In this case, the flux and mass ratio are uniquely related to only the size ratio.

It appears that the Arecibo system was not resolved from the \textit{Gaia} observations (the projected separation in the AL direction is always significantly smaller than the resolution unit). \new{As a result, the semi-major axis is not constrained. To further fit this parameter, }we included into the fit the two relative position data points measured from ground-based stellar occultation \citep{2022MPBu...49....3G}.  The target function is now 
\begin{equation}
    \ln L = -\frac{1}{2}\left[\sum{\left[\frac{\bar{x}-\delta\eta}{\sigma_{\bar{x}}}\right]^2} + \sum\left[\frac{l-l_c}{\sigma_l}\right]^2 + \sum  \left[\frac{pa-pa_c}{\sigma_{pa}}\right]^2 \right]
,\end{equation} where $l$ is the separation and $pa$ the position angle, with their associated uncertainties.

\section{Results}

\begin{figure}
   \centering
   \includegraphics[width=\linewidth]{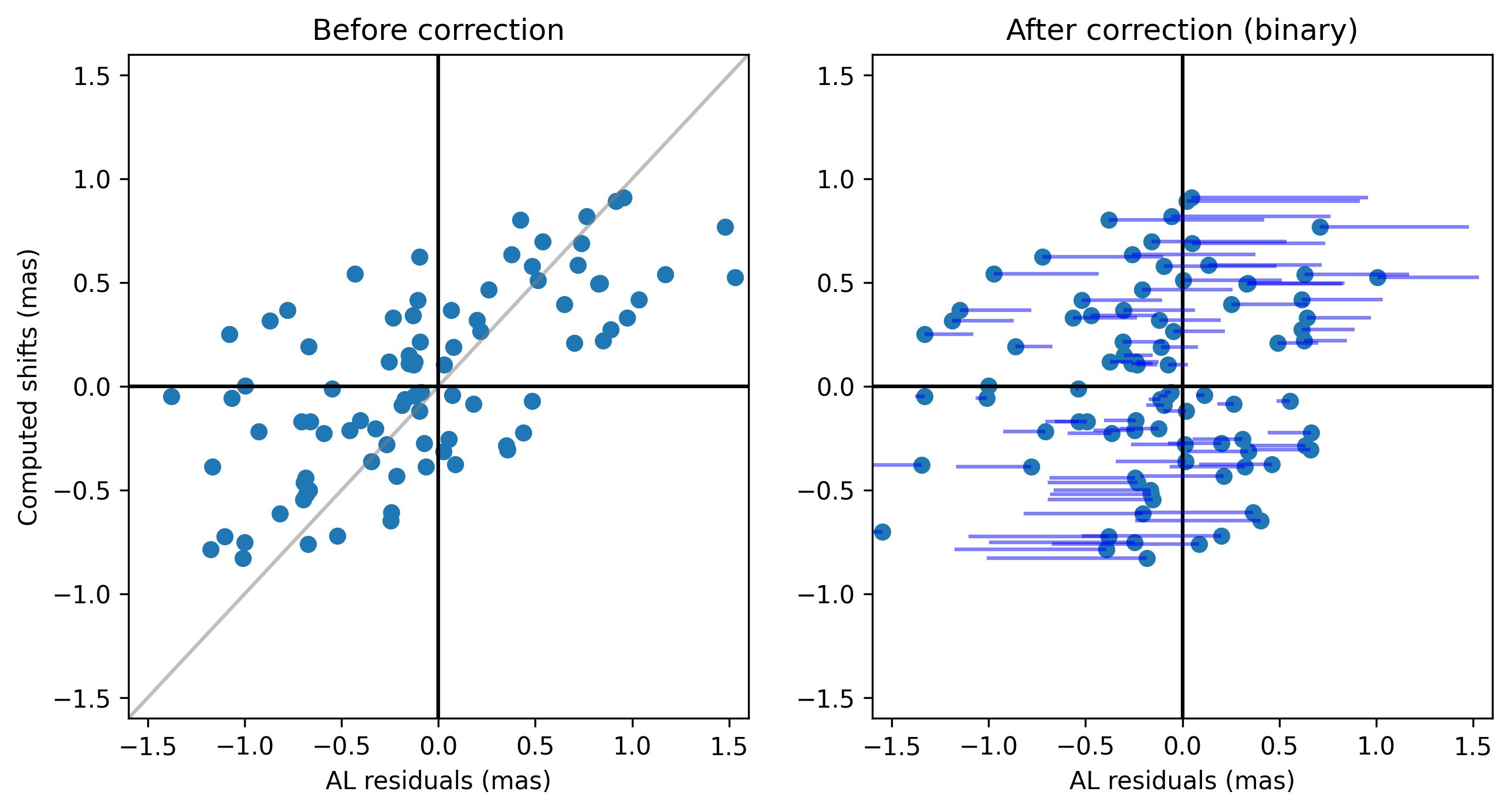}
    \caption{Comparison between the average heliocentric transit residuals and the computed binary photocentre--barycentre offsets. The right panel shows the equivalent plot after correcting for the computed shift. The horizontal lines represent the displacement between the data points before and after correction. A typical error bar, not shown in the figure, is of the order of $0.6\,$mas.}
   \label{fig:corr}
\end{figure}

We obtain a Gaussian-like posterior probability distribution function in the orbital parameters; the fitted orbital parameters are given in Table~\ref{tab:res}. The left panel of Fig.~\ref{fig:corr} shows the computed photocentre shifts from the binary model as a function of the AL residuals from NIMA, which has a Pearson correlation coefficient of 0.67.  On the right, the residuals are displayed after correcting for the binary astrometric wobble. The goodness of fit is thus globally improved: the RMS of the post-fit is decreased by approximately 25\%. The total mass of the system,  obtained from the third Kepler law,  is well constrained by $a$ and $T$, resulting in $M=4.3 \pm 0.5 \times 10^{15}\ $kg. We note that the orbital period is very well defined, thanks to the long time-frame of the FPR data. The precision on the total mass hence remains  mostly limited by the knowledge of the semi-major axis. 

\new{In addition, two solutions of the spin pole and period of the primary are derived from the \textit{Gaia} photometric data \citep{durech23}. We  include them in Table~\ref{tab:res} with their uncertainties. The similarity between the orbital pole and one of the spin poles and the corresponding spin period, together with the relatively small eccentricity, suggests a tidally locked system on a quasi-circular equatorial prograde orbit.}

\begin{figure}
    \centering
    \includegraphics[width=\linewidth]{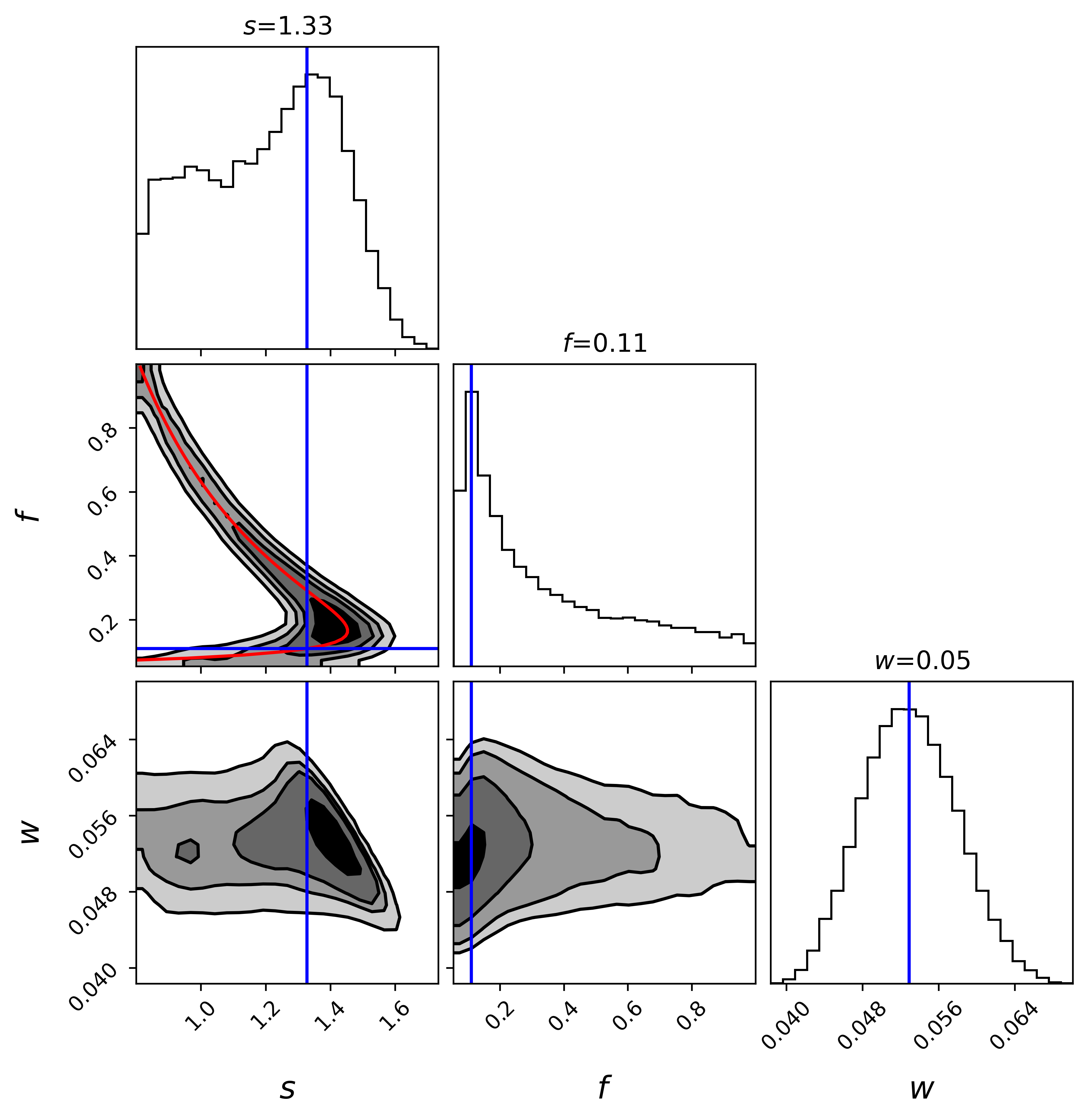}
    \caption{Posterior distribution of the parameters $s$, $f$, and $w$. The astrometric wobble amplitude $w$ is computed from the equation introduced in the text. The red line is plotted solving the relationship of $f$ and $s$ with $w = 0.053$ (median of its distribution). The blue line is taken at the mode of the distribution of $f = 0.11$, and we take it as one of the most probable solutions of this system. The corresponding value of $s$ is $1.33$.}
    \label{fig:paras}
\end{figure}

The corner plot of all estimated parameters is given in Appendix~\ref{ap:dist}. \new{This figure shows that the inclusion of occultation data, and the relative positions, is essential for deriving the semi-major axis and also narrowing down the other orbital parameters.} Moreover, one can determine that there is no correlation between $s$, $f$, and the other parameters. We show in Fig.~\ref{fig:paras} the posterior distribution of $s$ and $f$, with the global (unresolved) astrometric wobble amplitude $w = f/(1+f)-q/(1+q)$. One can see that $w$ is well constrained, while the flux ratio and the correcting factor are correlated (through mass ratio). 

\begin{table}[h]
    \centering
    \caption{Fitted and derived parameters for the relative orbit of the Arecibo binary.}
    \label{tab:res}
    \begin{tabular}{l c|c}
    \toprule
    Parameters && Value \\
    \midrule
     \underline{Orbital parameters}\\
        semi-major axis  [km] &$a$&   $46.8^{+1.8}_{-2.0}$\\ 
        \\[-0.8em]
        eccentricity &$e$ &$0.06^{+0.04}_{-0.04}$\\
        \\[-0.8em]
        longitude of ascending node [deg] &$\Omega$& $357^{+6}_{-6}$\\
        \\[-0.8em]
        inclination [deg] &$i$&$27^{+3}_{-3}$\\
        \\[-0.8em]
        argument of pericentre [deg] &$\omega$& $233^{+44}_{-44}$\\
        \\[-0.8em]
        pericentre passed time\tablefootmark{a} [day] & $t_0$&$0.39^{+0.16}_{-0.16}$\\
        \\[-0.8em]
        period [h] &$T$ &$32.9741^{+0.0004}_{-0.0004}$\\
        \\[-0.8em]
    
        \midrule
       \new{ \underline{Reference parameters \tablefootmark{b} }}\\
        \new{ 1st spin pole solution [deg]} &\new{$[\lambda,\beta]$} & \new{$ [269^{+3}_{-3}, 62^{+2}_{-2}]$}\\
        \\[-0.8em]
        \new{2nd spin pole solution [deg] }&\new{$[\lambda,\beta]$} &\new{$ [71^{+3}_{-3}, 63^{+2}_{-2}]$}\\
        \\[-0.8em]
        \new{spin period [h]} &$\new{P}$ & \new{32.974 }\\
        \\[-0.8em]

    \midrule
      \underline{Derived parameters} \\
        \\[-0.8em]
      longitude of orbital pole [deg] & $\lambda$ & $267^{+6}_{-6}$\\
      \\[-0.8em]
      latitude  of orbital pole [deg] & $\beta$   & $ 63^{+3}_{-3}$\\
      \\[-0.8em]
      flux ratio            & $f$ & 0.11 \tablefootmark{c} \\
      \\[-0.8em]
      mass ratio            & $q$ & 0.05 \tablefootmark{c}\\
      \\[-0.8em]
     wobble amplitude \tablefootmark{e} [mas] & $w$ & $0.053^{+0.005}_{-0.005}$\\
     \\[-0.8em]
      total mass [$10^{15}$\, kg]     & $M$ & $4.3^{+0.5}_{-0.5}$\\
      \\[-0.8em]
      primary mass [$10^{15}$\, kg]   & $M_1$ & $4.1$ \tablefootmark{c} \\
      \\[-0.8em]
      secondary mass [$10^{15}$\, kg] & $M_2$ & $0.2$ \tablefootmark{c}\\
      \\[-0.8em]
      primary density [kg/m$^3$]   & $\rho_1$ & $1200$ \tablefootmark{d} \\
      \\[-0.8em]
      secondary density [kg/m$^3$] & $\rho_2$ & $1600$ \tablefootmark{d}\\
\bottomrule
 \end{tabular}
 \tablefoot{The pole and parameters are given in the J2000 ecliptic frame; the error bars correspond to the 16th and 84th percentiles.\\
\tablefoottext{a}{The reference epoch is JD 2456877.70264 (UTC).}
\tablefoottext{b}{\new{Solutions from \citet{durech23}. The uncertainty of the rotation period P is of the order of the last decimal place.}}
\tablefoottext{c}{The results here are drawn from the posterior distribution of $f$ and $s$; see Sect.~\ref{S:disc}.}
\tablefoottext{d}{Considering an effective size from WISE.}
\tablefoottext{e}{Unsolved astrometric wobble amplitude.}
}
 \end{table}

\section{Discussion}
\label{S:disc}

Several assumptions are made in this preliminary work: The photocentre offsets $O_p$ in Eq.~(\ref{eq:r}), due to the phase effect of each component, were neglected; we   checked that they remain smaller than the astrometric wobble $\delta\eta$. 
Similarly, possible offsets in the relative position from the occultation data  modelling were neglected. 
The orbit computations were performed sequentially, starting with  the heliocentric orbit for the centre of mass. We did not consider any precession of the binary orbit over the five\,years of \textit{Gaia} observations due to external perturbations. We also did not consider  a possible dynamical flattening of the primary; assuming a theoretical $c/a$ ratio as high as 1/2, the mass and density determination could be underestimated by approximately 15\%. All these effects remain small, but will be included in a future work.

The relative separation of the Arecibo binary on the AL axis is small, and the system is never resolved by the \textit{Gaia} telescope. All Arecibo data fell into the first two cases described in Eq.~(\ref{eq:cases}), which are almost identical given the small separation $\Delta\eta$ \citep{LL:LL-136}.
As seen in Fig.~\ref{fig:paras}, the astrometric wobble amplitude is converging to $w=0.053\pm0.005$, while $f$, $s$, and hence $q$ are correlated. The correlation curve for $w=0.053$ is plotted as a red line in the figure with flux ratio range from 0 to 1. 
If we take the mode of $f$ as the most probable solution we have $f = 0.11$, $s = 1.33$, and hence $q=0.05$. It should be noted that the correcting factor $s$ differs significantly from unity, which suggests a non-spherical shape,  a different density, or a different albedo, or all three combined. 

The thermal infrared survey WISE \citep{wise19} gives a surface equivalent diameter $D_{eff}=19.7\pm0.2$\,km for Arecibo. If we consider a pseudo system composed of two spheres of diameter $D_{1}$ and $D_{2}$, and same albedo, we obtain from $f =( D_{2}^2/D_1^2)$ the absolute sizes $D_{1}=18.7$\,km and $D_{2}=6.2$\,km. Combining further with the mass ratio $q$, a bulk density of each component is derived, resulting in $\rho_1 = 1200\,$kg/m$^3$ and $\rho_2$ = $1600\,$kg/m$^3$. These density values are in line with the fact that Arecibo, together with (90) Antiope and (379) Huenna, is a member of the Themis family, an icy family in the outer belt \citep{2010GeoRL..3710202C, 2015Icar..254..150H, 2016A&A...586A..15M}. Furthermore, irrespective of the components' absolute sizes, but still assuming similar albedos, the density ratio is $\rho_2/\rho_1=s\approx1.33$. Thus, the secondary is more dense, or less porous than the primary. In general, for two spheres, $(\rho_2/\rho_1).(A_1/A_2)^{3/2}=s$, so that a more reflective primary would lower this density ratio.
In any case, the shapes and sizes of the components appear to be poorly known, with inconsistent values (by a factor of 1.5) from either occultation data or other radiometric data from Akari. For this reason, no error bars are given here for the density estimates.

\section{Conclusion}

By analysing \textit{Gaia} data from the FPR release, completed by two ground-based occultations, we  derived the orbital and physical parameters of the binary (4337) Arecibo from an analysis of its relative orbit. 

\new{In previous studies, for  example \citet{margot2002}, \citet{Ostro2006}, \citet{Naidu2012}, and the summary in \citet{margot2015}, radar observations have proved to be a powerful method for detecting and inferring the individual physical properties of near-Earth binary systems. In our work, we presented another method for obtaining such results for a small main-belt binary, by analysing high-precision optical astrometric data.} The procedure can be applied to other \new{known} systems observed as unresolved and resolved binaries by \textit{Gaia} and ground- or space-based telescopes. \new{Moreover, it can be implemented on the binary candidates from \citet{Liberato2024}, and can further provide the ephemeris of the mutual orbit and predicted position of the secondary. We will also include perturbed two-body problem for the dynamical modelling in future work.}


The accurate estimation of the bulk density of each component is complicated by the limited knowledge of their sizes and shapes (and hence volume). More observations are needed to constrain their absolute or relative sizes. This can be done with additional occultations, and more chords distributed along the two bodies. Several occultations in the coming years can be of interest to this purpose. For instance, there are two stellar occultation predictions for the system in December 2024 and January 2025 (see Fig.~\ref{ap:occult}).
It is also possible to predict the seasons of mutual phenomena (occultations, eclipses) in such a binary system \citep{emelyanov21}. 
The next season in this system will occur in November 2025 (see Fig.~\ref{ap:events}), while at solar elongation of approximately 60$^{\circ}$, with a predicted magnitude drop of $\Delta G \approx 0.1$ in the photometric lightcurve. 
Such mutual phenomena or stellar occultation observations will be very valuable for better constraining the components sizes, and hence the respective density.
Additional data at longer elongations, or with resolved observations, would help in constraining the orbit, and  better separating the  mass and flux ratios.

\begin{acknowledgements}
The authors would like to thank F. Arenou (Paris observatory) for fruitful discussions.\\
This work has made use of data from the European Space Agency (ESA) mission \textit{Gaia} (\url{https://www.cosmos.esa.int/gaia}), processed by the \textit{Gaia} Data Processing and Analysis Consortium (DPAC, \url{https://www.cosmos.esa.int/web/gaia/dpac/consortium}). Funding for the DPAC has been provided by national institutions, in particular the institutions participating in the \textit{Gaia} Multilateral Agreement. We made use of the {\tt emcee} ~\citep{emcee} and {\tt corner} ~\citep{corner} python package for data fitting and visualization. In addition, we made use of ephemerides from IMCCE \url{https://ssp.imcce.fr}.
\end{acknowledgements}

\bibliographystyle{aa} 
\bibliography{ref} 

\begin{thebibliography}{20}
\expandafter\ifx\csname natexlab\endcsname\relax\def\natexlab#1{#1}\fi

\bibitem[{{Castillo-Rogez} \& {Schmidt}(2010)}]{2010GeoRL..3710202C}
{Castillo-Rogez}, J.~C. \& {Schmidt}, B.~E. 2010, \grl, 37, L10202

\bibitem[{{Desmars}(2015)}]{2015A&A...575A..53D}
{Desmars}, J. 2015, \aap, 575, A53

\bibitem[{{Emelyanov} {et~al.}(2023){Emelyanov}, {Kovalev}, \& {Varfolomeev}}]{emelyanov21}
{Emelyanov}, N.~V., {Kovalev}, M.~Y., \& {Varfolomeev}, M.~I. 2023, \mnras, 522, 165

\bibitem[{Foreman-Mackey(2016)}]{corner}
Foreman-Mackey, D. 2016, The Journal of Open Source Software, 1, 24

\bibitem[{Foreman-Mackey {et~al.}(2013)Foreman-Mackey, Hogg, Lang, \& Goodman}]{emcee}
Foreman-Mackey, D., Hogg, D.~W., Lang, D., \& Goodman, J. 2013, Publications of the Astronomical Society of the Pacific, 125, 306–312

\bibitem[{{Gaia Collaboration} {et~al.}(2023){Gaia Collaboration}, {David}, {Mignard}, {Hestroffer}, {Tanga}, {Spoto}, {Berthier}, {Pauwels}, {Roux}, {Barbier}, {Cellino}, {Carry}, {Delbo}, {Dell'Oro}, {Fouron}, {Galluccio}, {Klioner}, {Mary}, {Muinonen}, {Ordenovic}, {Oreshina-Slezak}, {Panem}, {Petit}, {Portell}, {Brown}, {Thuillot}, {Vallenari}, {Prusti}, {de Bruijne}, {Arenou}, {Babusiaux}, {Biermann}, {Creevey}, {Ducourant}, {Evans}, {Eyer}, {Guerra}, {Hutton}, {Jordi}, {Lammers}, {Lindegren}, {Luri}, {Randich}, {Sartoretti}, {Smiljanic}, {Walton}, {Bailer-Jones}, {Bastian}, {Cropper}, {Drimmel}, {Katz}, {Soubiran}, {van Leeuwen}, {Audard}, {Bakker}, {Blomme}, {Casta{\~n}eda}, {De Angeli}, {Fabricius}, {Fouesneau}, {Fr{\'e}mat}, {Guerrier}, {Masana}, {Messineo}, {Nicolas}, {Nienartowicz}, {Pailler}, {Panuzzo}, {Riclet}, {Seabroke}, {Sordo}, {Th{\'e}venin}, {Gracia-Abril}, {Teyssier}, {Altmann}, {Benson}, {Burgess}, {Busonero}, {Busso}, {C{\'a}novas}, {Cheek}, {Clementini}, {Damerdji}, {Davidson}, {de
  Teodoro}, {Delchambre}, {Fraile Garcia}, {Garabato}, {Garc{\'\i}a-Lario}, {Garralda Torres}, {Gavras}, {Haigron}, {Hambly}, {Harrison}, {Hatzidimitriou}, {Hern{\'a}ndez}, {Hodgkin}, {Holl}, {Jamal}, {Jordan}, {Krone-Martins}, {Lanzafame}, {L{\"o}ffler}, {Lorca}, {Marchal}, {Marrese}, {Moitinho}, {Nu{\~n}ez Campos}, {Osborne}, {Pancino}, {Recio-Blanco}, {Riello}, {Rimoldini}, {Robin}, {Roegiers}, {Sarro}, {Schultheis}, {Siopis}, {Smith}, {Sozzetti}, {Utrilla}, {van Leeuwen}, {Weingrill}, {Abbas}, {{\'A}brah{\'a}m}, {Abreu Aramburu}, {Aerts}, {Altavilla}, {{\'A}lvarez}, {Alves}, {Anderson}, {Antoja}, {Baines}, {Baker}, {Balog}, {Barache}, {Barbato}, {Barros}, {Barstow}, {Bartolom{\'e}}, {Bashi}, {Bauchet}, {Baudeau}, {Becciani}, {Bedin}, {Bellas-Velidis}, {Bellazzini}, {Beordo}, {Berihuete}, {Bernet}, {Bertolotto}, {Bertone}, {Bianchi}, {Binnenfeld}, {Blazere}, {Boch}, {Bombrun}, {Bouquillon}, {Bragaglia}, {Braine}, {Bramante}, {Breedt}, {Bressan}, {Brouillet}, {Brugaletta}, {Bucciarelli}, {Butkevich},
  {Buzzi}, {Caffau}, {Cancelliere}, {Cannizzo}, {Carballo}, {Carlucci}, {Carnerero}, {Carrasco}, {Carretero}, {Carton}, {Casamiquela}, {Castellani}, {Castro-Ginard}, {Cesare}, {Charlot}, {Chemin}, {Chiaramida}, {Chiavassa}, {Chornay}, {Collins}, {Contursi}, {Cooper}, {Cornez}, {Crosta}, {Crowley}, {Dafonte}, {de Laverny}, {De Luise}, {De March}, {de Souza}, {de Torres}, {del Peloso}, {Delgado}, {Dharmawardena}, {Diakite}, {Diener}, {Distefano}, {Dolding}, {Dsilva}, {Dur{\'a}n}, {Enke}, {Esquej}, {Fabre}, {Fabrizio}, {Faigler}, {Fatovi{\'c}}, {Fedorets}, {Fern{\'a}ndez-Hern{\'a}ndez}, {Fernique}, {Figueras}, {Fournier}, {Gai}, {Galinier}, {Garcia-Gutierrez}, {Garc{\'\i}a-Torres}, {Garofalo}, {Gerlach}, {Geyer}, {Giacobbe}, {Gilmore}, {Girona}, {Giuffrida}, {Gomel}, {Gomez}, {Gonz{\'a}lez-N{\'u}{\~n}ez}, {Gonz{\'a}lez-Santamar{\'\i}a}, {Gosset}, {Granvik}, {Gregori Barrera}, {Guti{\'e}rrez-S{\'a}nchez}, {Haywood}, {Helmer}, {Helmi}, {Henares}, {Hidalgo}, {Hilger}, {Hobbs}, {Hottier}, {Huckle},
  {Jab{\l}o{\'n}ska}, {Jansen}, {Jim{\'e}nez-Arranz}, {Juaristi Campillo}, {Khanna}, {Kordopatis}, {K{\'o}sp{\'a}l}, {Kostrzewa-Rutkowska}, {Kun}, {Lambert}, {Lanza}, {Le Campion}, {Lebreton}, {Lebzelter}, {Leccia}, {Lecoeur-Taibi}, {Lecoutre}, {Liao}, {Liberato}, {Licata}, {Lindstr{\o}m}, {Lister}, {Livanou}, {Lobel}, {Loup}, {Mahy}, {Mann}, {Manteiga}, {Marchant}, {Marconi}, {Mar{\'\i}n Pina}, {Marinoni}, {Marshall}, {Mart{\'\i}n Lozano}, {Mart{\'\i}n-Fleitas}, {Marton}, {Masip}, {Massari}, {Mastrobuono-Battisti}, {Mazeh}, {McMillan}, {Meichsner}, {Messina}, {Michalik}, {Millar}, {Mints}, {Molina}, {Molinaro}, {Moln{\'a}r}, {Monari}, {Mongui{\'o}}, {Montegriffo}, {Montero}, {Mor}, {Mora}, {Morbidelli}, {Morel}, {Morris}, {Mowlavi}, {Munoz}, {Muraveva}, {Murphy}, {Musella}, {Nagy}, {Nieto}, {Noval}, {Ogden}, {Pagani}, {Pagano}, {Palaversa}, {Palicio}, {Pallas-Quintela}, {Panahi}, {Payne-Wardenaar}, {Pegoraro}, {Penttil{\"a}}, {Pesciullesi}, {Piersimoni}, {Pinamonti}, {Pineau}, {Plachy}, {Plum}, {Poggio},
  {Pourbaix}, {Pr{\v{s}}a}, {Pulone}, {Racero}, {Rainer}, {Raiteri}, {Ramos}, {Ramos-Lerate}, {Ratajczak}, {Re Fiorentin}, {Regibo}, {Reyl{\'e}}, {Ripepi}, {Riva}, {Rix}, {Rixon}, {Robichon}, {Robin}, {Romero-G{\'o}mez}, {Rowell}, {Royer}, {Ruz Mieres}, {Rybicki}, {Sadowski}, {S{\'a}ez N{\'u}{\~n}ez}, {Sagrist{\`a} Sell{\'e}s}, {Sahlmann}, {Sanchez Gimenez}, {Sanna}, {Santove{\~n}a}, {Sarasso}, {Sarrate Riera}, {Sciacca}, {Segovia}, {S{\'e}gransan}, {Shahaf}, {Siebert}, {Siltala}, {Slezak}, {Smart}, {Snaith}, {Solano}, {Solitro}, {Souami}, {Souchay}, {Spina}, {Spitoni}, {Squillante}, {Steele}, {Steidelm{\"u}ller}, {Surdej}, {Szabados}, {Taris}, {Taylor}, {Teixeira}, {Tisani{\'c}}, {Tolomei}, {Torra}, {Torralba Elipe}, {Trabucchi}, {Tsantaki}, {Ulla}, {Unger}, {Vanel}, {Vecchiato}, {Vicente}, {Voutsinas}, {Weiler}, {Wyrzykowski}, {Zhao}, {Zorec}, {Zwitter}, {Balaguer-N{\'u}{\~n}ez}, {Leclerc}, {Morgenthaler}, {Robert}, \& {Zucker}}]{david23}
{Gaia Collaboration}, {David}, P., {Mignard}, F., {et~al.} 2023, \aap, 680, A37

\bibitem[{{Gault} {et~al.}(2022){Gault}, {Nosworthy}, {Nolthenius}, {Bender}, \& {Herald}}]{2022MPBu...49....3G}
{Gault}, D., {Nosworthy}, P., {Nolthenius}, R., {Bender}, K., \& {Herald}, D. 2022, Minor Planet Bulletin, 49, 3

\bibitem[{{Hargrove} {et~al.}(2015){Hargrove}, {Emery}, {Campins}, \& {Kelley}}]{2015Icar..254..150H}
{Hargrove}, K.~D., {Emery}, J.~P., {Campins}, H., \& {Kelley}, M. S.~P. 2015, \icarus, 254, 150

\bibitem[{{Liberato} {et~al.}(2024){Liberato}, {Tanga}, {Mary}, {Minker}, {Carry}, {Spoto}, {Bartczak}, {Sicardy}, {Oszkiewicz}, \& {Desmars}}]{Liberato2024}
{Liberato}, L., {Tanga}, P., {Mary}, D., {et~al.} 2024, arXiv e-prints, arXiv:2406.07195

\bibitem[{Lindegren(2022)}]{LL:LL-136}
Lindegren, L. 2022

\bibitem[{{Mainzer} {et~al.}(2019){Mainzer}, {Bauer}, {Cutri}, {Grav}, {Kramer}, {Masiero}, {Sonnett}, \& {Wright}}]{wise19}
{Mainzer}, A.~K., {Bauer}, J.~M., {Cutri}, R.~M., {et~al.} 2019, {NEOWISE Diameters and Albedos V2.0}, NASA Planetary Data System, urn:nasa:pds:neowise\_diameters\_albedos::2.0

\bibitem[{{Margot} {et~al.}(2002){Margot}, {Nolan}, {Benner}, {Ostro}, {Jurgens}, {Giorgini}, {Slade}, \& {Campbell}}]{margot2002}
{Margot}, J.~L., {Nolan}, M.~C., {Benner}, L.~A.~M., {et~al.} 2002, Science, 296, 1445

\bibitem[{{Margot} {et~al.}(2015){Margot}, {Pravec}, {Taylor}, {Carry}, \& {Jacobson}}]{margot2015}
{Margot}, J.~L., {Pravec}, P., {Taylor}, P., {Carry}, B., \& {Jacobson}, S. 2015, in Asteroids IV, 355--374

\bibitem[{{Marsset} {et~al.}(2016){Marsset}, {Vernazza}, {Birlan}, {DeMeo}, {Binzel}, {Dumas}, {Milli}, \& {Popescu}}]{2016A&A...586A..15M}
{Marsset}, M., {Vernazza}, P., {Birlan}, M., {et~al.} 2016, \aap, 586, A15

\bibitem[{{Muinonen} {et~al.}(2022){Muinonen}, {Berthier}, {Cellino}, {David}, {De Angeli}, {Delb{\'o}}, {Dell-Oro}, {Galluccio}, {Hestroffer}, {Mignard}, {Pauwels}, {Spoto}, \& {Tanga}}]{gdr3-22}
{Muinonen}, K., {Berthier}, J., {Cellino}, A., {et~al.} 2022, {Gaia DR3 documentation Chapter 8: Solar System Objects}, Gaia DR3 documentation, European Space Agency; Gaia Data Processing and Analysis Consortium. Online at <A href=``https://gea.esac.esa.int/archive/documentation/GDR3/index.html''>https://gea.esac.esa.int/archive/documentation/GDR3/index.html</A>, id. 8

\bibitem[{{Naidu} {et~al.}(2012){Naidu}, {Margot}, {Busch}, {Taylor}, {Nolan}, {Howell}, {Giorgini}, {Benner}, {Brozovic}, \& {Magri}}]{Naidu2012}
{Naidu}, S.~P., {Margot}, J.~L., {Busch}, M.~W., {et~al.} 2012, in AAS/Division of Dynamical Astronomy Meeting, Vol.~43, AAS/Division of Dynamical Astronomy Meeting \#43, 7.07

\bibitem[{{Ostro} {et~al.}(2006){Ostro}, {Margot}, {Benner}, {Giorgini}, {Scheeres}, {Fahnestock}, {Broschart}, {Bellerose}, {Nolan}, {Magri}, {Pravec}, {Scheirich}, {Rose}, {Jurgens}, {De Jong}, \& {Suzuki}}]{Ostro2006}
{Ostro}, S.~J., {Margot}, J.-L., {Benner}, L. A.~M., {et~al.} 2006, Science, 314, 1276

\bibitem[{{Pravec} \& {Scheirich}(2012)}]{2012P&SS...73...56P}
{Pravec}, P. \& {Scheirich}, P. 2012, \planss, 73, 56

\bibitem[{{Tanga} {et~al.}(2023){Tanga}, {Pauwels}, {Mignard}, {Muinonen}, {Cellino}, {David}, {Hestroffer}, {Spoto}, {Berthier}, {Guiraud}, {Roux}, {Carry}, {Delbo}, {Dell'Oro}, {Fouron}, {Galluccio}, {Jonckheere}, {Klioner}, {Lefustec}, {Liberato}, {Ord{\'e}novic}, {Oreshina-Slezak}, {Penttil{\"a}}, {Pailler}, {Panem}, {Petit}, {Portell}, {Poujoulet}, {Thuillot}, {Van Hemelryck}, {Burlacu}, {Lasne}, \& {Managau}}]{tanga23}
{Tanga}, P., {Pauwels}, T., {Mignard}, F., {et~al.} 2023, \aap, 674, A12

\bibitem[{{{\v{D}}urech} \& {Hanu{\v{s}}}(2023)}]{durech23}
{{\v{D}}urech}, J. \& {Hanu{\v{s}}}, J. 2023, \aap, 675, A24

\end{thebibliography}

\begin{appendix}\label{appendix}
\onecolumn

\section{Full results}\label{ap:dist}

\begin{figure}[h]
    \centering
    \includegraphics[width=\linewidth]{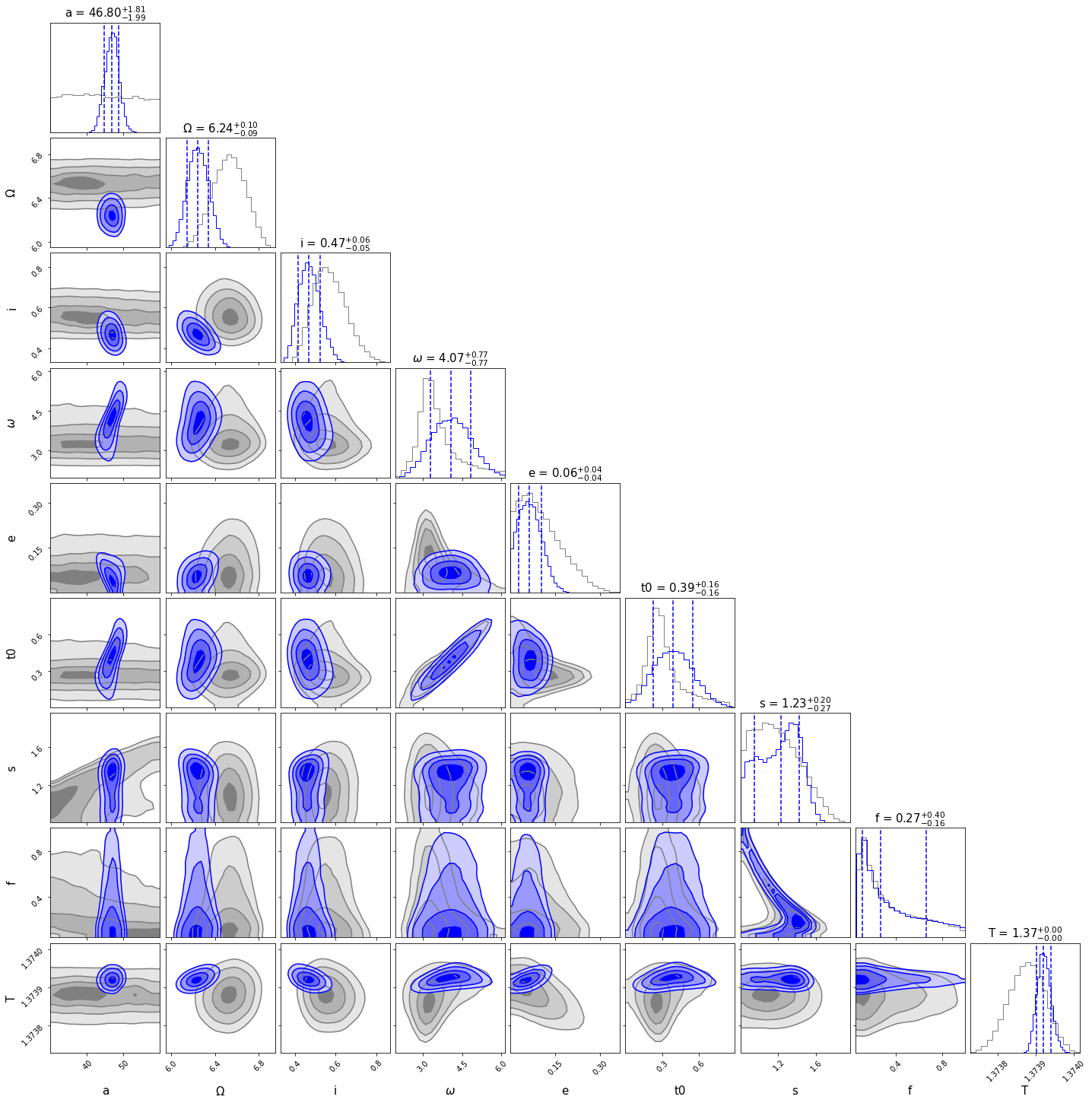}
    \caption{Posterior distribution of all nine fitted  parameters. All angles are in radians, in ecliptic J2000 reference frame; the semi-major axis is in kilometres; and the time-related parameters are in days. \new{In grey: MCMC sample fit from \textit{Gaia} data only; in blue: MCMC samples, including the data from the two occultations. The text and dashed line are given at the 16th, 50th, and 84th percentiles for the distribution including occultation.}}
    \label{fig:res_occ}
\end{figure}

\begin{figure}[h]
    \centering
    \includegraphics[width=0.68\linewidth]{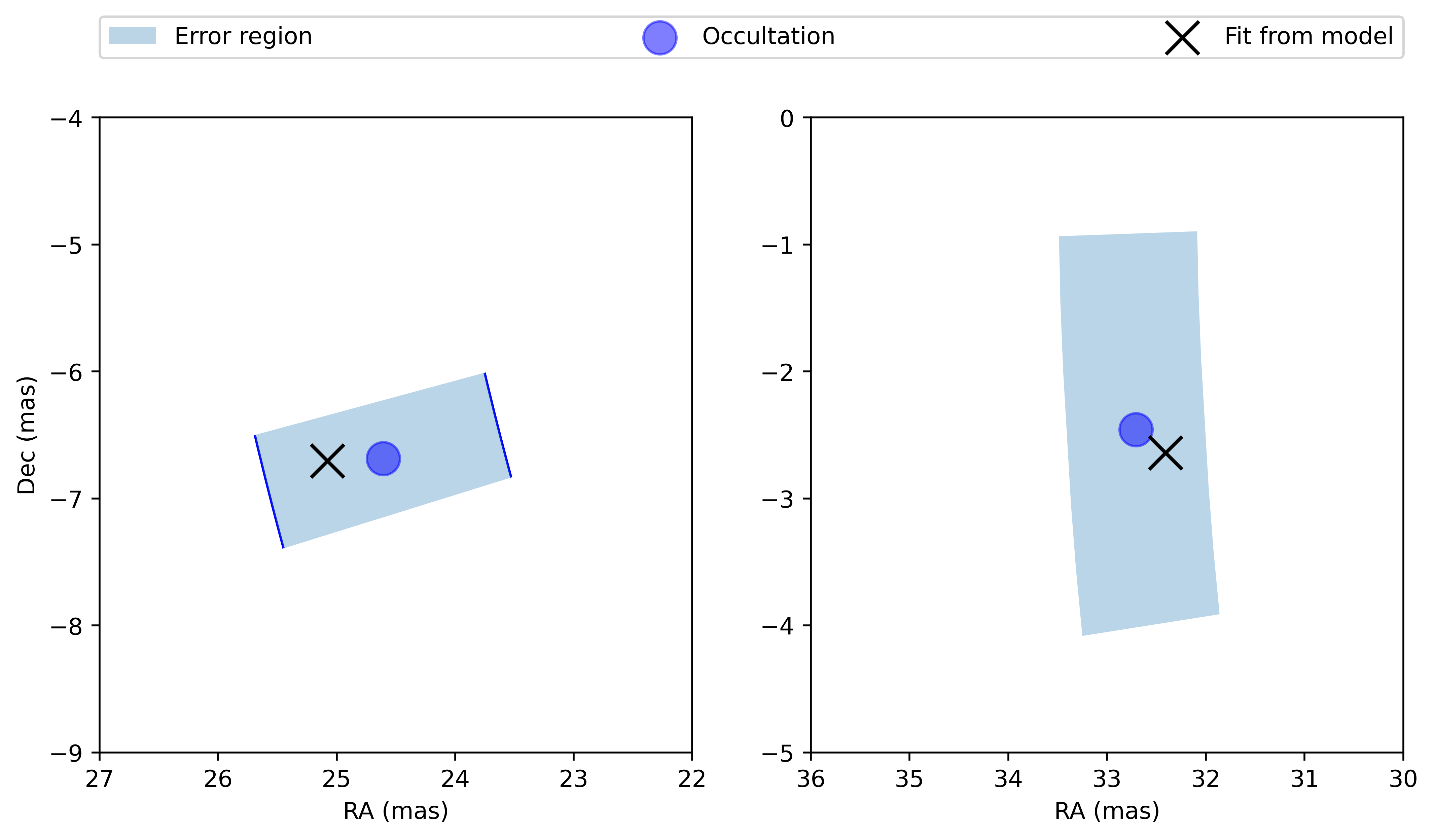}
    \caption{Residuals of the satellite position between the two observed occultations (blue dot) and the prediction (black cross). Left: 19 May 2021; Right: 9 June 2021. The primary's position is at the centre (0,0). The shaded area is plotted following the error bar in \citet{2022MPBu...49....3G}.}
    \label{fig:res_occ}
\end{figure}

\newpage
\section{Event predictions}\label{ap:events}

\subsection{Stellar occultations}\label{ap:occult}


\begin{figure}[h]
    \centering
           \begin{subfigure}{0.45\linewidth}
            \centering
                \includegraphics[width=\linewidth]{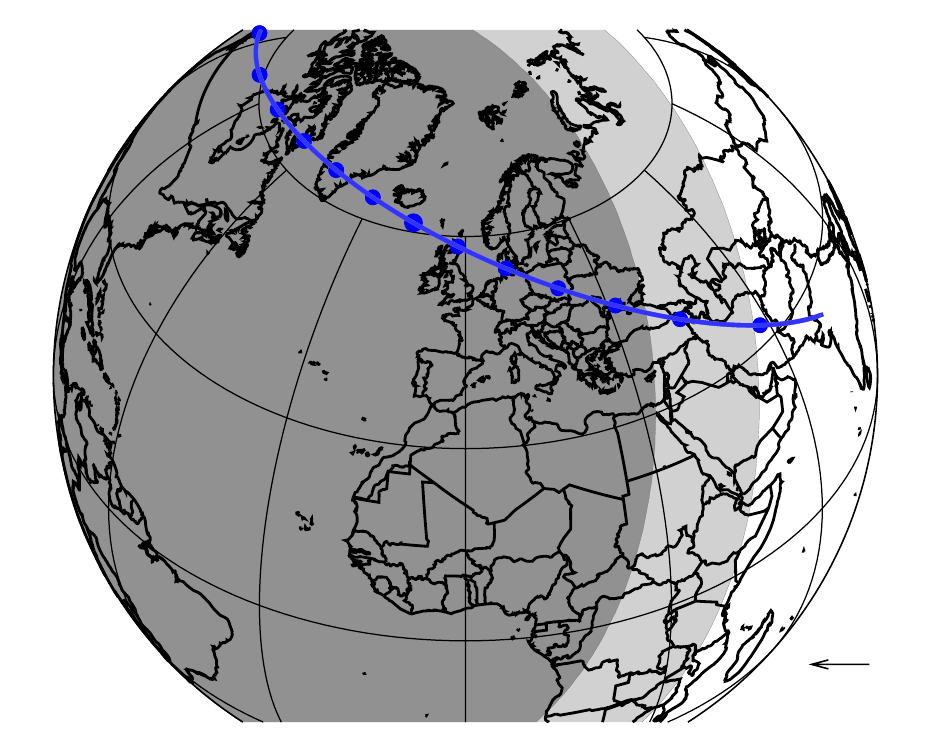}
           \end{subfigure}
    \hfill
           \begin{subfigure}{0.45\linewidth}
        \centering
                \includegraphics[width=\linewidth]{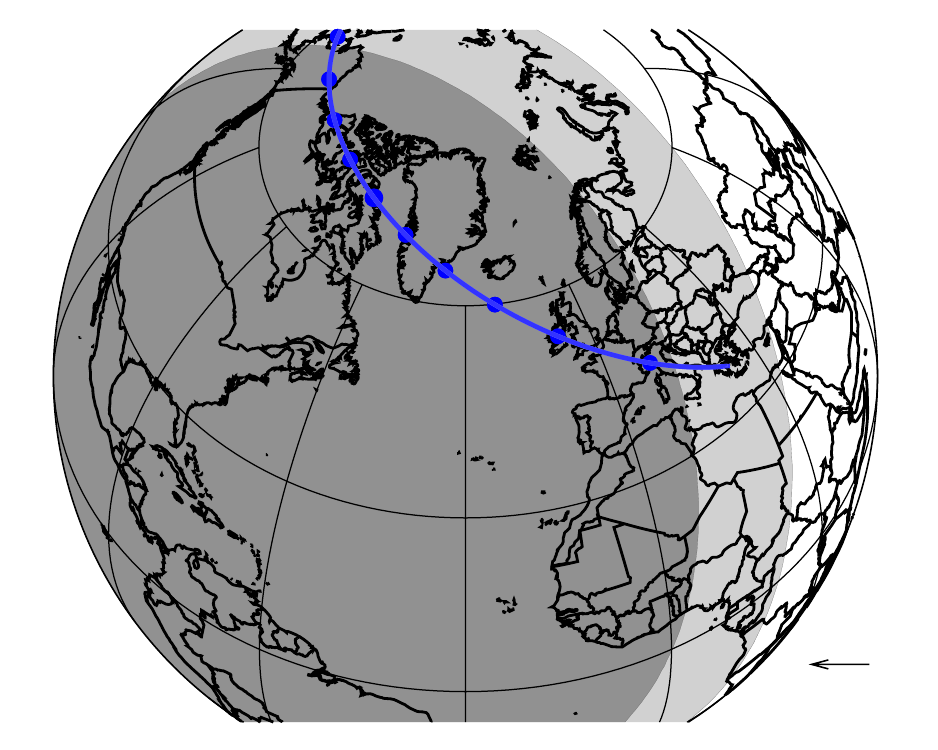}
            \end{subfigure}
        \caption{Two stellar occultation predictions of Arecibo system in the near future. Left: Event on 21 December 2024, with the star magnitude of $13.3$ and a drop in magnitude of $2.5$. Right: Event on 18 January 2025, with the star magnitude of $13.1$ and a drop in magnitude of $2.5$.}
        \label{fig:occ}
\end{figure}
\newpage
\subsection{Mutual events}\label{ap:mutualevents}
\begin{figure}[h]
    \centering
    \includegraphics[width=\linewidth]{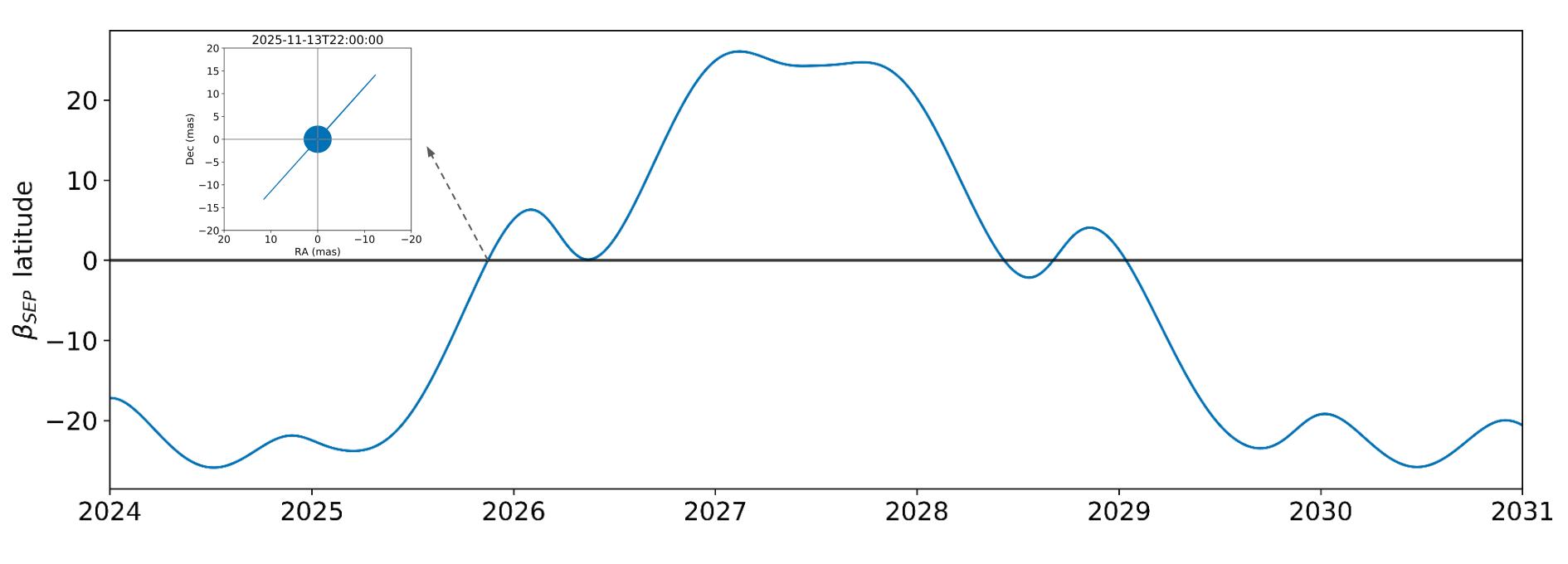}
    \caption{Prediction of the Earth’s elevation (sub-Earth point latitude $\beta_{SEP}$) above the orbital plane. The zoomed-in plot shows the plane-of-sky view at the time of one predicted mutual event; the blue line is the projected relative orbit. }
    \label{fig:mutual}
\end{figure}

\end{appendix}

\end{document}